\documentclass[aps,prb,preprint,floatfix,superscriptaddress,showpacs,amssymb,amsmath]{revtex4}
\usepackage{graphicx}

\begin{document}

\title{Hydrodynamic effects in interacting Fermi electron jets}
\author{Alexander~O.~Govorov}
\affiliation{Department of Physics and Astronomy, and The
Nanoscale and Quantum Phenomena Institute, Ohio University,
Athens, Ohio 45701}
\author{Jean~J.~Heremans}
\affiliation{Department of Physics and Astronomy, and The
Nanoscale and Quantum Phenomena Institute, Ohio University,
Athens, Ohio 45701}

\date{\today }

\begin{abstract}

We theoretically study hydrodynamic phenomena originating from
electron-electron collisions in a two-dimensional Fermi system. We
demonstrate that an electron beam sweeping past an aperture
creates a pumping effect, attracting carriers from this aperture.
This pumping effect originates from the specific electric
potential distribution induced by the injected electrons. In the
regions nearby the main stream of injected electrons, a positive
potential is induced by the injected electrons. Thus, the normally
repulsive Coulomb interaction leads to an attractive force in the
Fermi system. This quantum pumping mechanism in a Fermi system
differs qualitatively from the Bernoulli pumping effect in
classical liquids. We also discuss possible experimental
realizations.
\end{abstract}


\maketitle




Electron-electron (e-e) scattering in a degenerate electron gas
forms one of the central concepts of Fermi-liquid theory
\cite{Landau,Landau1}. Electron states with energy slightly above
the Fermi surface are not stationary, since they decay accompanied
by the creation of electron-hole pairs in the Fermi sea
\cite{Landau}. The total momentum in the e-e scattering process is
conserved, and hence the e-e collisions do not in the main affect
the conductivity. At the same time, the e-e collisions play an
important role in experiments which involve the phase coherence
time of the electron \cite{Webb-Willett,mag-imp}, and in physical
effects implicating hydrodynamics
\cite{hydro,hydro1,Axel,Govorov}. At low temperatures, the e-e
collision processes slow down due to Fermi statistics
\cite{Chaplik,Quinn,AA}, leading, in the ideal two dimensional
electron system (2DES), to the inverse e-e lifetime rapidly
decreasing upon lowering the temperature $T$ \cite{Chaplik,Quinn}:

\begin{equation}
\frac{1}{\tau_{ee}}=\frac{(k_BT)^2}{h
E_F}[\ln{\frac{E_F}{k_BT}}+\ln{\frac{4}{a_0^*k_F}}+1],
\label{tauee}
\end{equation}
where $h$, $E_F$, and $v_F$ are the Planck's constant, the Fermi
energy, and the Fermi velocity, respectively; $a_0^*$ denotes the
effective Bohr radius. To render hydrodynamic effects in 2DESs
observable, e-e collisions should occur more often than impurity
scattering events. Mathematically, this condition is obvious:
$l_{ee}<l_p$, where $l_{ee}=v_F\tau_{ee}$ is the electron mean
free path related to e-e scattering and $l_p$ is the momentum
relaxation mean free path. The above condition is satisfied in
typical high-mobility GaAs/GaAlAs heterostructures, in the
temperature range $T\sim 5-35~K$ (Fig.~1a) \cite{ref11}.
Therefore, the hydrodynamic effects should be accessible in
experimental low-temperature studies. Here we describe a new
hydrodynamic phenomenon in a low-temperature electron plasma which
occurs when an electron beam is injected from a narrow aperture
into a 2DES (Fig.~1b). We show that the injected beam creates an
unusual potential pattern which can result in a hydrodynamic
pumping effect. This pumping effect exists in mesoscopic
structures where the injected electrons experience just few e-e
collisions. In contrast to the Bernoulli effect in classical
liquids, the predicted quantum pumping behaves linearly in the
excitation, and can only be observed in a Fermi system at low
temperatures.

Another potential application of the predicted hydrodynamic effect
concerns measurements of the electron phase coherence time,
$\tau_{\phi}$. The low-temperature behavior of $\tau_{\phi}$ in
nanostructures is a long standing problem. According to the
standard theory, e-e scattering becomes very weak at low
temperatures and therefore $\tau_{\phi}$ should tend to infinity
at zero temperature.  However, the experiments demonstrate that
$\tau_{\phi}$ saturates to a finite value at low temperatures
\cite{Webb-Willett,mag-imp}. Magnetic-impurity scattering has been
implicated amongst others \cite{mag-imp}. Yet, e-e scattering
itself is a nontrivial problem in mesoscopic structures, where an
electron experiences scattering by boundaries as well as
impurities.  For example, it is accepted that in diffusive metals,
the formula $1/\tau_{ee}\propto T^2$ should be strongly modified
due to disorder \cite{AA}. Measurements of the hydrodynamic effect
predicted in this work can directly reveal the contribution of e-e
scattering to the phase coherence time in mesoscopic ballistic
structures at low temperatures.

To describe hydrodynamic effects in a stationary electron beam
(Fig.~1b) we employ the linearized Boltzmann equation:

\begin{equation}
{\bf v}\frac{\partial f_1}{\partial {\bf r}}+e{\bf
E}\frac{\partial f_0}{\partial {\bf p}}=J(f_1), \label{BE}
\end{equation}
where $f_1({\bf r},{\bf p})$ is the distribution function of the
non-equilibrium electrons, ${\bf r}=(x,y)$ and ${\bf p}$ are the
two-dimensional coordinate and momentum, respectively;
$f_0(\epsilon)$ is the equilibrium Fermi function, ${\bf v}={\bf
p}/m$ is the electron velocity, and $\epsilon$ and $m$ are the
energy and the effective mass of the electron, respectively; ${\bf
E}$ is the in-plane electric field, $e=-|e|$ is the electron
charge, and $J$ is the collision integral. At low temperature, the
non-equilibrium electrons have energies close to the Fermi surface
and the function $f_1$ can be written as $f_1({\bf
r};\epsilon,\theta)=-\frac{\partial f_0}{\partial \epsilon}F({\bf
r};\theta)$, where $\theta$ is the angle between the velocity and
the direction $+x$, with $0<\theta<2\pi$. Our choice of $(x,y)$
coordinates is indicated in Fig. 1b.

We treat e-e collisions in the relaxation-time approximation. Due
to the complexity of the collision integral incorporating the
long-range Coulomb interaction, theoretical studies based on the
long-range Coulomb collision integral usually include only the
first e-e collision event \cite{hydro,Davies}. In contrast, the
relaxation-time approach allows us to obtain analytical results
for an infinite number of collisions and to describe fundamental
hydrodynamic effects appearing in the multi-collision regime.
Within the relaxation-time approximation, the collision integral
describing scattering events in the vicinity of the Fermi surface,
takes the form:

\begin{equation}
J(F)=-\frac{F-\overline{F}-2\cos{\theta}\overline{\cos{\theta}F}-
2\sin{\theta}\overline{\sin{{\theta}}F}}{\tau_{ee}}
\label{CI1}
\end{equation}
where $\bar{F}=\int_0^{2\pi}F(\theta)d\theta/2\pi$ and $\tau_{ee}$
is the e-e scattering time. Note that the e-e collision integral
(\ref{CI1}) conserves both the number of particles and the
momentum. In addition, the collision integral (\ref{CI1}) assumes
an ideal system without disorder and is valid if $l_{ee}<l_p$.


To solve the Boltzmann equation in the half-plane $x>0$ we need to
impose the boundary condition (BC) at the line $x=0$. Outside the
aperture, the BC describes the elastic collision of an electron
with an ideal border, $F(0,y;\theta)-F(0,y;\pi-\theta)=0$. In
order to express the injection process, we now introduce the
injection function $g(y,\theta)$, nonzero only in the angle
intervals $0<\theta<\pi/2$ and $(3/2)\pi<\theta<2\pi$. It is
convenient to consider the BC separately in two intervals,
$-\pi/2<\theta<\pi/2$ and $\pi/2<\theta<(3/2)\pi$. In the first
angular interval, the outgoing flux of electrons $v_FF(\theta)$
should equal the sum of the incoming and the injected fluxes,
$v_FF(\pi-\theta)+v_Fg(\theta)$. A similar argument is applied to
the second interval. In both intervals, the BC condenses to a
single formula:

\begin{equation}
F(0,y;\theta)-F(0,y;\pi-\theta)=g(y;\theta)-g(y;\pi-\theta).
\label{BC2}
\end{equation}
Importantly, the BC (\ref{BC2}) leaves some freedom in choosing
the function $F(0,y;\theta)$. A general solution of eq.~\ref{BC2}
can be written as $F(0,y;\theta)=w_1(y;\theta)+w_2(y;\theta)$,
where the function $w_1(\theta)$ is determined by the injection
$g(y;\theta)$, whereas the function $w_2(\theta)$ is arbitrary and
satisfies the elastic-collision condition
$w_2(y;\theta)-w_2(y;\pi-\theta)=0$.

The BCs in this problem are given by the equation \ref{BC2} at the
line $x=0$ and by the condition
$F(x\rightarrow\infty)\rightarrow0$ at infinity. To solve this
problem in a rather convenient way, we now consider a symmetric
problem, involving the entire two-dimensional plane and two
symmetric current sources (two jets) in the vicinity of ${\bf
r}=0$. Including the two current sources, we write the linearized
Boltzmann equation as follows:

\begin{equation}
{\bf v}\frac{\partial F}{\partial {\bf r}}-e{\bf E}{\bf
v}=J(F)+G(y;\theta)\delta(x)+ G(y;\pi-\theta)\delta(x),
\label{BE2}
\end{equation}
where the function $G(y;\theta)$ describes the injection. By
exploiting the symmetry of the problem and integrating
eq.~\ref{BE2} over $x$ in the vicinity of $x=0$, we determine that
the solution of eq.~\ref{BE2} satisfies the necessary BC
(\ref{BC2}) with $G(\theta)=v_F\cos{\theta}g(\theta)$. Equation
(\ref{BE2}) can be solved using Fourier transformation,

\begin{equation}
i{\bf v}{\bf k}F_k-e{\bf E}_k{\bf v}=-\frac{F_k-\overline{F}_k-
2\cos{\theta}\overline{\cos{\theta}F}_k-
2\sin{\theta}\overline{\sin{{\theta}}F}_k}{\tau_{ee}}+G_{k_y}(\theta)+
G_{k_y}(\pi-\theta), \label{BEk}
\end{equation}
where $F_{\bf k}(\theta)=\int_{-\infty}^{+\infty} d^2{\bf
r}e^{-i{\bf k}{\bf r}}F({\bf r};\theta)$. The electric field ${\bf
E}$ in eq.~\ref{BEk} originates from the non-equilibrium
electrons. We now assume that the heterostructure containing the
2DES is covered by a top metallic gate, with the distance between
the 2DES and the top gate $d$ smaller than any lateral size of the
system.  This condition does not unduly restrict experimental
situations.  Imposing this condition ensures that the potential
$\phi$ in the plane of 2D plasma is proportional to the
non-equilibrium 2DES density $\delta n({\bf r})$. Thus, we can
write $\phi({\bf r})=(4\pi ed/\epsilon_s)\delta n({\bf r})$, where
$\epsilon_s$ denotes the dielectric constant of the semiconductor
\cite{Govorov,Frank}. After Fourier transformation, we obtain
${\bf E}_k=-i{\bf k}\phi_k=-i{\bf k}(4\pi ed/\epsilon_s)\delta
n_k$. Simultaneously, the electron density is expressed through
the function $\overline{F}$, $\delta n_k=\bar{F}_kD_{2D}$, where
$D_{2D}=m/\pi\hbar^2$ represents the 2D density of states. By
manipulating eq.~\ref{BEk} and integrating over angles, we obtain
a closed system of equations for $\overline{F}_k$,
$\overline{\cos{\theta}F}_k$, and $\overline{\sin{\theta}F}_k$.
The 2DES density can be determined by the functions

\begin{equation} \bar{F}_{\bf
k}=\frac{I_1\frac{2k_{0}}{k^2}(k_0W_0-1)+I_2}
{(1+4d/a^*_0)(1-k_0W_0)}, \label{Fk}
\end{equation}
\begin{equation}
W_0(k)=\frac{1}{\sqrt{k^2+k_0^2}}, \ \
I_1=\int_0^{2\pi}G_{k_y}(\theta)\frac{d\theta}{\pi v_F}, \ \
I_2({\bf k})=\int_0^{2\pi}
\frac{G_{k_y}(\theta)+G_{k_y}(\pi-\theta)}{ik\cos(\theta-\alpha)+k_0}\frac{d\theta}{2\pi
v_F},
\end{equation}
where $\tan{\alpha}=k_y/k_x$ and
$k_{0}=(v_F\tau_{ee})^{-1}=1/l_{ee}$.
We also derive analytical expressions for the current densities,
$j_x({\bf k})=v_FD_{2D}\overline{\cos{\theta}F}_k$ and $j_y({\bf
k})=v_FD_{2D}\overline{\sin{\theta}F}_k$.

For the injection function, we choose

\begin{equation}
G(y,\theta)=G_{max}e^{-\frac{y^2}{L^2}}
\frac{\Theta(\theta_0-|\theta|)}{2\theta_0}, \label{G}
\end{equation}
with $\Theta(\theta)=1$ if $\theta>0$ and $0$ otherwise. The
parameters $L$ and $\theta_0$ describe the spatial width of the
aperture and the width of the angular distribution of injected
electrons, respectively.  The parameter $G_{max}$ is proportional
to the injected current density in the middle of the aperture.
Using the injection function, $G_{max}$ can be related to the
total injected current, by $I=|e|D_{2D}L/(2\sqrt{\pi})G_{max}$.
Typically, the resistance of the injecting aperture exceeds the
resistance of the leads and thus the potential drop across the
aperture, $\Delta V$, dominates the potential applied to the
system. Using the Landauer-B\"uttiker formalism, we can write
$I=(2e^2/h)N\Delta V$ \cite{Conductance}, where $N$ represents the
number of conducting modes in the aperture.  Although quantization
of the conductance is not necessary for the observation of the
hydrodynamic effect ($N$ can be large), the Landauer-B\"uttiker
formalism forms a convenient framework to link the experimental
variables, $L$, $N$, $\Delta V$ and $I$.  By combining the above
equations, we obtain: $G_{max}=4\sqrt{\pi}|e|N/(h LD_{2D})\Delta
V$.  In addition, it was assumed that the injected electrons
possess energies in the vicinity of the Fermi surface.

Numerical results for the non-equilibrium density distribution
follow from the reverse transformation, $\delta n=D_{2D}\int
exp(i{\bf k}{\bf r})\bar{F}_k d{\bf k}/(2\pi)^2$.  Figure 2 shows
that $\delta n$ is large and positive inside the ballistic beam,
as expected.  But $\delta n$ is depressed to negative values in
regions nearby the main stream of injected electrons. The negative
$\delta n$ corresponds to depletion. In Fig.~3, we schematically
show the streamlines associated with injection. Interestingly, in
the regions adjacent to the main stream, the currents flow towards
the beam.  Qualitatively, we can understand this behavior in terms
of e-e scattering: the injected electrons create an effective
pressure and scatter the background Fermi-sea electrons toward the
right. Numerical calculations show that this happens not in the
main stream, where the density of excess electrons is high, but
rather on the sides of the main stream. The currents flowing
toward the main stream tend to compensate for the lack of
electrons.

The electron depletion in the vicinity of the injected beam can
lead to carrier pumping toward this beam.  The potential induced
by the non-equilibrium electrons can be expressed as $\phi({\bf
r})=(4\pi ed/\epsilon_s)\delta n({\bf r})$, reflecting the fact
that a local net charge will lead to a local potential of the same
sign. If a detector aperture is situated in the region of electron
depletion ($\delta n < 0$), the detector will experience a
positive potential (Fig.3).  In voltage measurement mode, the
detector can be considered a closed reservoir with net zero
current.  A positive potential denotes a lack of electrons in the
detector reservoir, the result of a net pumping of electrons
toward the injected beam.  We can also consider the pumping in
terms of currents.  Suppose that we turn on the injected current
$I$ at $t=0$. In the region of the detector window, the pumping
effect extracts electrons from the detector. After some time, the
system reaches steady state, and the net current through the
detector window vanishes. Hence a counter current of electrons
into the detector must be generated, by a positive potential on
the detector lead. If on the contrary current is allowed to flow,
electrons will be pumped through the aperture in steady state.  To
achieve the steady state pumping, we can connect the detector
contact to the Fermi sea in the right-hand side, including a
resistor to maintain the current at a level sufficiently low for
the latter to be regarded as a weak perturbation.  In all above
cases, the positive potential induced by the hydrodynamic effects
in the vicinity of the injected beam, effectively result in a
pumping phenomenon.

According to eq.~\ref{tauee}, the length $l_{ee}=v_F\tau_{ee}$
strongly depends on temperature, leading to a temperature
dependence of the pumping effect. In the ideal disorderless
system, the pumping potential depends on the ratios $L/l_{ee}$,
$x/l_{ee}$, and $y/l_{ee}$ (Fig. 1b). It is revealing to analyze
the temperature dependence of the induced electric potential at a
fixed position with coordinates $(x_A,y_A)$ where a detector
window can be located (Fig. 3). Calculations (Fig.~3) indicate
that the potential induced by the injected beam at the point $A$
with coordinates $(x_A=2~\mu m, y_A=-2~\mu m)$ is negative at low
and high temperatures, while for intermediate temperatures,
$3-25~K$, the hydrodynamic effects render the potential positive.
In this regime, $l_{ee}\sim x_A\sim y_A$ and electron-electron
interaction leads to just a few scattering events in the vicinity
of the detector. Therefore, the pumping effect occurs only in the
regime of few scattering events. Our calculations show that the
pumping voltage increases by decreasing the angle $\theta_0$.
Hence, better experimental conditions to observe the pumping
signal can be found in relatively wide apertures with flared
openings, where the collimation effect is typically stronger
\cite{Molenkamp}. In addition, since $\phi\propto\delta n\propto
G_{max}\propto\Delta V\propto I$, the pumping voltage is a linear
function of the applied voltage or injected current.

In the case of one-dimensional injection ($L\rightarrow\infty$),
the function $\delta n(x)$  can be obtained from eq.~\ref{Fk} by
setting $G_{k_y}\propto\delta(k_y)$. The calculated electron
density clearly demonstrates the hydrodynamic effect: the Fermi
sea depletes in the region $x\sim l_{ee}$ due to the effective
pressure created by the injected beam (Fig.~4). This effect can
also be applied toward a pumping mechanism involving a
quasi-one-dimensional channel featuring an injecting barrier and a
detector window, as illustrated in Fig.~4.

It is interesting to compare our results with the hydrodynamics of
classical liquids. Of course, the local non-linear equations of
fluid mechanics differ qualitatively from the non-local kinetic
Boltzmann equation. For example, the Bernoulli pumping effect,
resulting from the spatial variation of local speed in the moving
liquid, is quadratic in terms of fluid speed, hence non-linear. As
result the Bernoulli pumping effect remains unchanged if we
reverse the current. In the 2DES case discussed in this work, the
pumping force is linear in terms of applied voltage or current and
changes sign as we reverse the current flow. The linearity is a
consequence of the electron Fermi statistics. Indeed, at low
temperatures, we can consider the flow equally in terms of
electrons or of holes near the Fermi surface. Mathematically, the
described linear response originates naturally from the linearized
Boltzmann equation. Physically, when the negative voltage is
applied to the injecting contact, the injected beam consists of
electrons which induce an effective pumping force for the
electrons from a detector (Fig.~3). When a positive voltage is
applied, the injected beam can be treated as a beam of holes near
the Fermi surface; these holes now induce an effective pumping
force for the holes from the detector and therefore the pumping
voltage reverses its sign, along with the applied voltage.  The
concurrent voltage reversals lead to a linear dependence at low
excitations.  The linear dependence stands in contrast to
classical liquids where currents consist of like particles and the
above Fermi-gas description cannot be applied.

To conclude, we have shown that the Coulomb interaction in an
injected ballistic beam results in attractive forces which can be
exploited toward a pumping effect. The quantum pumping described
by us is a peculiar property of a degenerate 2DES and is
qualitatively different from the hydrodynamic effects in classical
liquids.

We thank Axel Lorke for helpful discussions. This work was
supported by NSF (DMR-0094055) and by the Condensed Matter and
Surface Science Program at Ohio University.

\newpage

\begin{figure}
\includegraphics[width=2.30in,angle=-90]{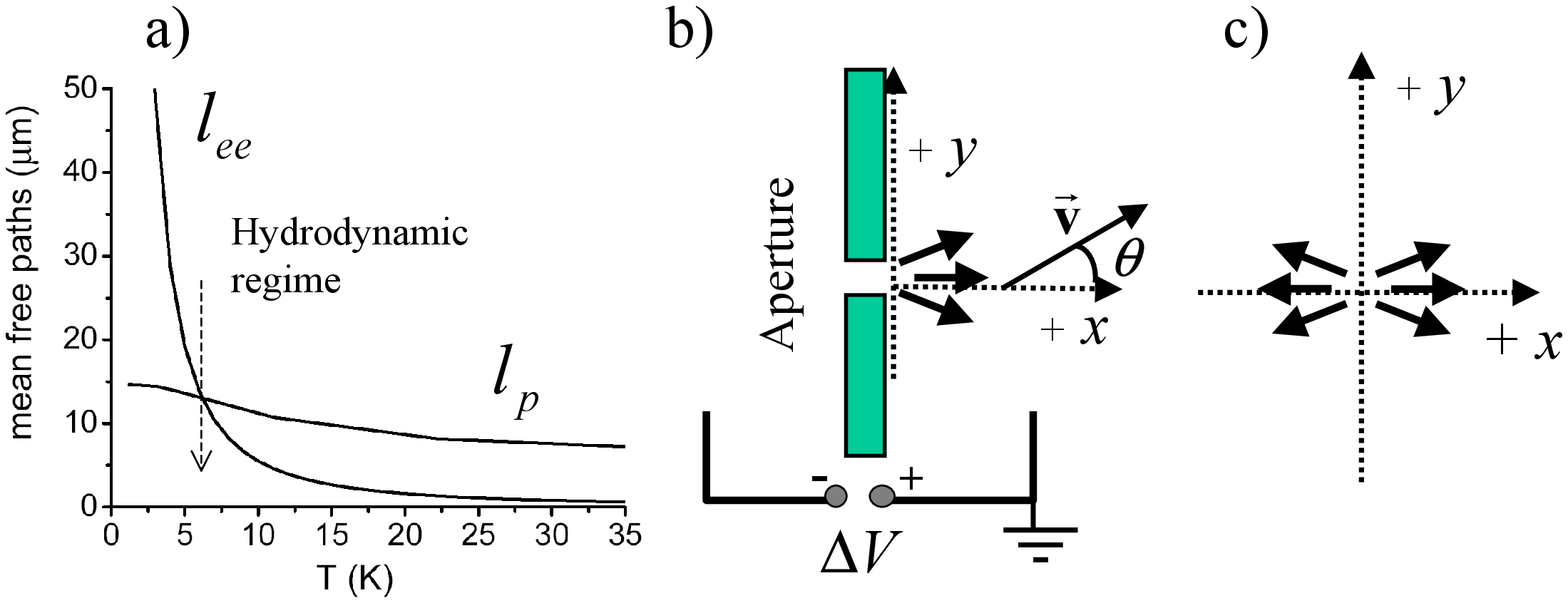}
\caption{\ \\ \ \label{fig1} \ \\ \ (a) Momentum relaxation and
electron-electron mean free paths, $l_p$ and $l_{ee}$, as a
function of temperature. The parameter $l_p$ is obtained from
mobility measurements on a GaAs/AlGaAs heterostructure at the 2DES
density $4.8*10^{11}~cm^{-2}$ [13]. $l_{ee}$ is calculated using
eq.~14 from ref.~10. (b) Geometry of the system with an injected
electron beam. (c) Schematic of two symmetric jets in a
two-dimensional plasma. \ \\ \ \\ \ \\ \ }
\end{figure}

\newpage

\begin{figure}
\includegraphics[width=3.50in,angle=-90]{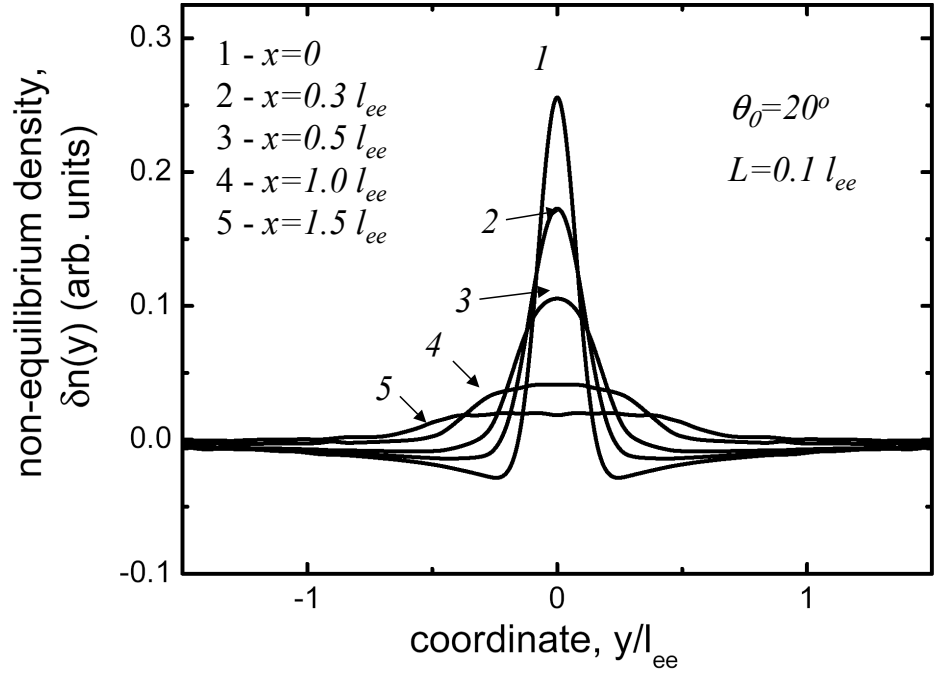}
\caption{\ \\ \ \label{fig2} \ \\ \ Calculated non-equilibrium
electron density as a function of the in-plane coordinates.}
\end{figure}

\newpage

\begin{figure}
\includegraphics[width=3.50in,angle=-90]{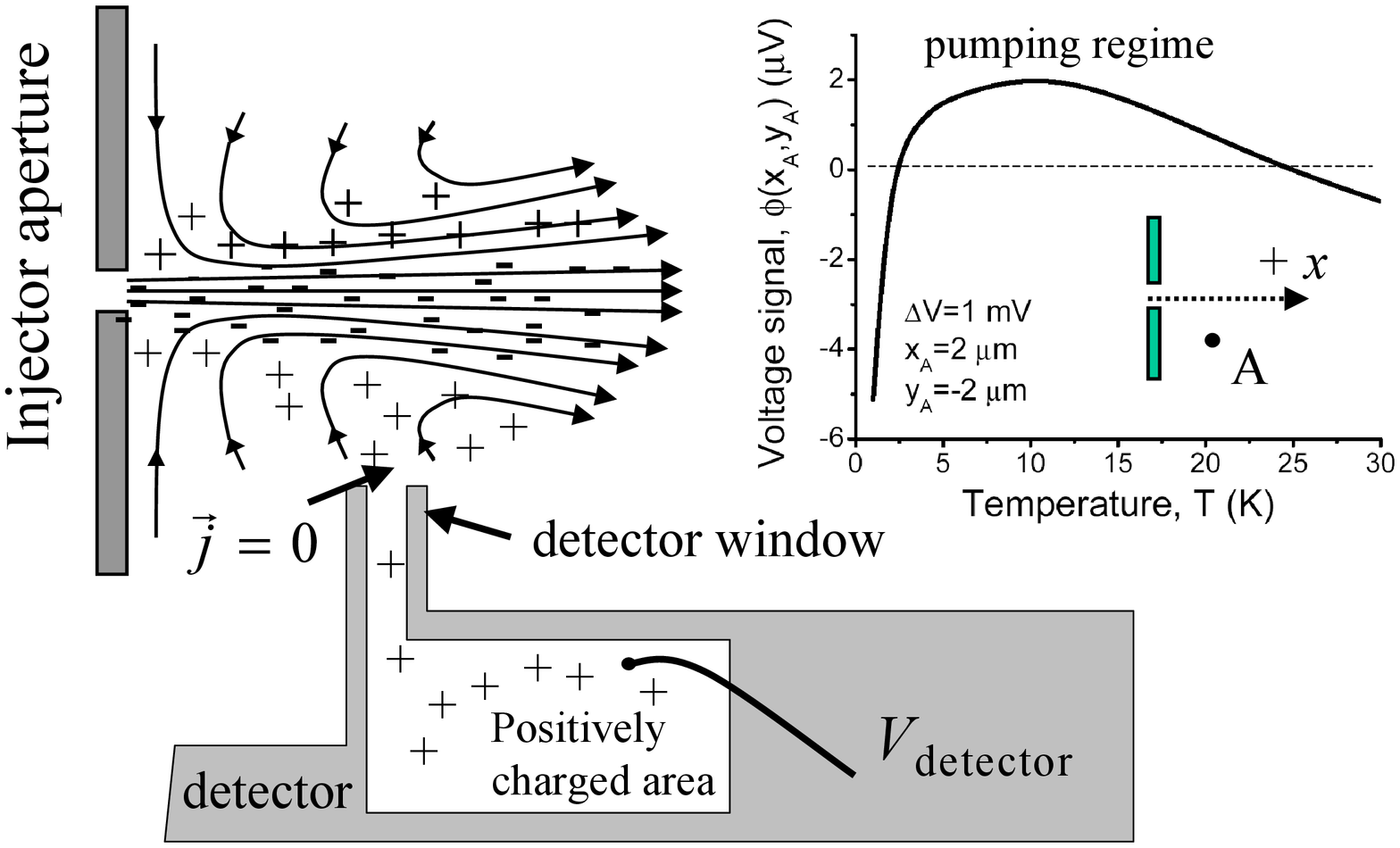}
\caption{\ \\ \ \label{fig3} \ \\ \ Schematic of a hydrodynamic
pump in a mesoscopically patterned 2DES. The stream lines and
density distribution are shown schematically. The detector serves
as a probe for the potential near the main beam of electrons.
Insert: The induced voltage at the point $A$ with coordinates
$(x_A,y_A)$ as a function of temperature; number of conducting
modes in the QPC, $N=10$; 2DES density = $4.8*10^{11}cm^{-2}$,
$d=400~\AA$, $a_0^*=100~\AA$, and $\epsilon=12.5$.}
\end{figure}

\newpage

\begin{figure}
\includegraphics[width=3.50in,angle=-90]{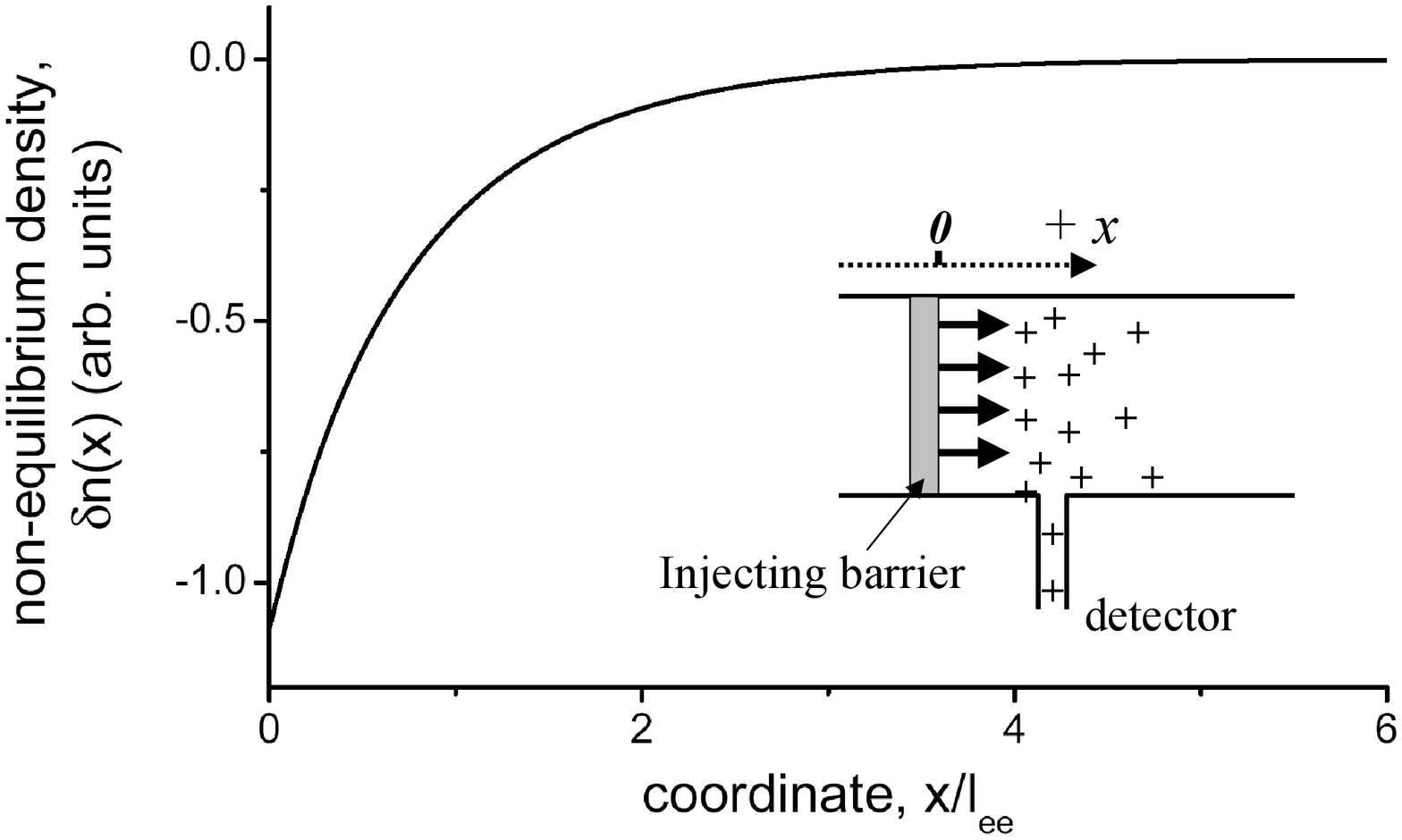}
\caption{\ \\ \ \label{fig4} \ \\ \ Calculated non-equilibrium
electron density as a function of the $x$-coordinate in the
one-dimensional system. Insert: Schematic of a hydrodynamic pump
utilizing a one-dimensional channel with an injecting barrier and
a detector aperture.}
\end{figure}

\end{document}